# A Natural Language Query Interface for Searching Personal Information on Smartwatches

Reza Rawassizadeh[1], Chelsea Dobbins[2], Manouchehr Nourizadeh[3], Zahra Ghamchili[4], Michael Pazzani[5]

[1]Department of Computer Science, Dartmouth College, US
[2]Department of Computer Science, Liverpool John Moore University, UK
[3] Vienna University of Technology, Austria
[4]Department of Education, University of Vienna, Austria
[5]Department of Computer Science and Engineering, University of California, Riverside, US

rrawassizadeh@acm.org, c.m.dobbins@ljmu.ac.uk, nourizadehbarabi.m.at@ieee.org, a0709907@univie.ac.at, michael.pazzani@ucr.edu

*Abstract*— **Currently, personal assistant systems, run on smartphones and use natural language interfaces. However, these systems rely mostly on the web for finding information. Mobile and wearable devices can collect an enormous amount of *contextual personal data* such as sleep and physical activities. These information objects and their applications are known as quantified-self, mobile health or personal informatics, and they can be used to provide a deeper insight into our behavior. To our knowledge, existing personal assistant systems do not support all types of quantified-self queries. In response to this, we have undertaken a user study to analyze a set of "textual questions/queries" that users have used to search their quantified-self or mobile health data. Through analyzing these questions, we have constructed a light-weight natural language based query interface - including a text parser algorithm and a user interface - to process the users' queries that have been used for searching quantified-self information. This query interface has been designed to operate on small devices, i.e. smartwatches, as well as augmenting the personal assistant systems by allowing them to process end users' natural language queries about their quantified-self data.**

## I. Introduction & Background

Recently, we have observed the proliferation of personal assistant applications, such as Apple's Siri[1], Google Google Now[2] and Microsoft Cortana[3]. Devices that provide personal assistant (PA) services, especially smartphones, are usually pervasive, but have small displays, which leads to limited interaction facilities [1]. Moreover, PA systems are connected to the cloud, and often dynamically update the variety of services that they provide. However, frequently changing a graphical user interface (GUI), based on the newly supported services is not trivial, and thus it is not practical to use GUIs for PA systems. Therefore, existing implementations of these systems rely on voice user interfaces, which parse text at the end. This makes natural language interfaces (NLI) one of the most important components in a PA system. There are NLIs that host GUIs but they have significantly less flexibility in comparison to textual NLIs [2].

In other words, NLIs overcomes limitations of interacting with small pervasive displays. However, ubiquitous devices that host PA systems are capable of collecting detailed contextual information about their users, such as sleep patterns, physical activities, communications, etc. Nevertheless, existing PA services do not benefit from utilizing all of the existing data that is available. Only simple contextual data, such as the user's current location, is often used, and other data objects will be read from the web [3].

We have conducted a user study on Amazon Mechanical Turk. We have collected 716 sample queries from 131 participants, who have smartphones[4]. These participants are familiar with PA systems, but they may not necessarily be familiar with Quantified Self (QS) systems. Employing participants who are not familiar with these systems is an advantage of our user study. Those participants who are unfamiliar with QS systems have helped us to identify queries, which have not been addressed by previous works, such as monitoring the usage of a cloth before washing it, or measuring the amount of radiation produced by smartphones.

Previous QS based user studies [4, 5, 6] provide promising results, but they focus only on users who are familiar with QS systems. In particular, Li et al. [4] have identified information that are important for users, including status, history, goals, discrepancies, context, and factors. Choe et al. [5] have focused on what QS users have learnt from collecting their data and how they perform the data collection. Oh et al. [6] have classified QS users and the issues that are associated with tracking. Rawassizadeh et al. [7] list challenges of collecting QS data and users' expectations from these systems. All of these works focus on identifying "motivation", "challenges" and "opportunities" that QS users faces, while using QS systems. We have benefited from their findings in framing our user study questions. However, our focus is on quantifying the QS users' request (queries) from the system.

---

[1] www.apple.com/ios/siri
[2] www.google.com/landing/now
[3] www.microsoft.com/en-us/mobile/experiences/cortana

[4] To allow full reproducibility of our results MTurk dataset, implementation of the algorithm with the query interface are all available. Please contact the first author to get access.

In this work, we propose a novel query interface for searching QS information. To implement this interface, first we need to identify elements of QS query construction. This means a major contribution of our research are NLI interfaces that identify queries [8,9,10,11].

Popescu et al. [8] have identified traceable questions and have proposed a holistic framework to convert natural language (NL) questions/queries into SQL commands. Other works [9,10] use Semantic Web ontologies for query construction. For instance, Querix [9] provides a model to extract general query syntax. This allows users to benefits from a clarification dialogue (GUI) that recommends correct sentences, based on their given sentence, to match it with their underlying ontology for information retrieval. NLP-Reduce [10] employs a simple domain-independent NLI interface to translate different user queries, i.e. keywords and sentence fragments to SPARQL queries. NaLIR [11] is a recent work that provides a holistic approach to convert human queries to SQL commands.

Nevertheless, in contrast to the listed NLI works, our work is not holistic and it covers only QS queries. Moreover, all of these works have a translation component from end users' questions to machine languages, such as SQL or SPARQL. However, small devices, such as smartwatches, have limited resources [12]. Therefore, a light resource-efficient customized query module is favored over SQL, SPARQL or any other resource intensive query engines, which is not trivial to implement on small devices.

Our contributions are two fold: first we conduct a study to understand QS queries of users to a PA system. Second, we propose a query interface, which includes an algorithm and a user interface that can identify important elements from user questions to construct machine understandable queries. It is notable that our work solely focuses only on a query interface, and thus retrieving information from underlying data stores is not in the scope of this paper.

## II. MATERIALS & METHODS

We have performed a survey using Amazon Mechanical Turk (MTurk). This study has created a very promising dataset of textual queries that illustrates what end-users are willing to search from their QS data. After collecting this data, we have applied thematic (qualitative) and linguistic analysis (quantitative analysis based on our qualitative results) on the queries to identify elements that constitute a successful query. Afterwards, we have introduced our algorithm and the smartwatch user interface to extract and understand the required elements from the given textual queries.

### A. User Study

*Objectives:* To implement a query interface that operates on personal data, we need to understand: (i) What types of knowledge are users seeking to find from their available personal data (on their mobile or wearable devices)? (ii) How do users query to search their desired information? To answer both questions, we have conducted a survey on MTurk.

*Participants:* For the user study, Bargas-Avila and Hornbæk [13] recommends to not only focus on users who are familiar with a system for a user study, but to also consider users who are unfamiliar with the system. Therefore, our participation criteria include participants who own smartphones, but they do not necessarily need to be familiar with QS systems. The survey results show that users, who were not familiar with QS technologies, propose several interesting queries. This insight enables better quantification of end user queries, and covers even queries for the sensors that do not existed yet, e.g. a sensor to determine if a cloth requires washing or not. We have surveyed 131 participants, including 71 males and 60 females, with an age range from 23 to 64 (mean=32, SD=8.19).

*Procedure:* Firstly, users were briefed with a one-page description about quantified-self technologiess [4,5,6,7]. This included information about what existing tools can track, and if the tracking tools require a manual user input or if the information can be collected automatically. For instance, "step count" can be collected automatically, whereas "valence of the mood" will need to be entered manually. To design the survey questions, we have used the method proposed by Rosenberg & Hovland [14] and thus our questions include cognitive, affective and behavioral lines of inquiry. In more technical sense, our survey includes three types of questions; the first type of questions focused on asking users "what" their expectation from a QS system is. This type of question was not limited to current capabilities of existing devices and let the user define their ideas for new applications. The second category of questions asked participants to describe "how" they are willing to search for the information. In contrast to the previous type, these questions were limited to existing technologies. Within this category, one specific question explicitly asks users to provide five to seven textual queries that they would use to search their personal information. Our linguistic analysis on constructing and modelling the query is based on these questions. The third category includes a single question (reverse score question). This question has three multiple choice sub-questions from the one-page description. It evaluates whether participants have understood the description and if their answer is valid for further analysis.

### B. Content Analysis and Query Construction

We have undertaken two types of analysis on the collected data from MTurk. Qualitative analysis has been undertaken using thematic analysis. Then, quantitative analysis has been conducted using linguistic analysis. Subsequently, based on both analysis, we construct the query parser and its user interface.

*1) Thematic Analysis*

Inductive thematic analysis has been applied on queries and other textual descriptions that participants have provided. A researcher has then examined this text and has identified themes.

Another researcher has then analyzed the results and a consensus has been reached on the majority of queries. Using Fleiss Kappa, the inter-rater reliability, results in k = 0.74. Based on Landis and Koch's interpretation [15], this is a substantial agreement between two researchers. Table 1. shows the categorization of queries based on "inductive thematic analysis".

Identified themes in Table 1 shows that there are limited categories of requests that a user could use to search their personal data repository. This finding has been used to constitute the foundation of our linguistic analysis algorithm.

**Table 1: Inductive thematic analysis of participants queries.**

| Health | eating, drinking, calorie monitoring, weight monitoring, medicine intake |
|---|---|
| Sleep | quality of sleep, quantity of sleep, wakeup time, wakeup mood |
| Behavior Control | eating behavior, game playing, social media activities, TV consumption, time management, stress |
| Sport | goals achievement, progress (endurance and performance) monitoring. |
| Financial Management | expenses (mobile data cost, online game purchases, etc.), grocery purchases (cost, nearby deals, discount items) |
| Memory and Forgetting | calendar events, appointments, bill payment, medicine intake, routine communications (calling parents, birthday reminder, etc.) |

*2) Linguistic Analysis*

To perform the quantitative analysis, we have stemmed the questions, removed punctuations and stop words, and have determined the frequency of the words in the collected queries.

Thematic analysis demonstrates the limitation of queries and their categories. The results illustrate that there are four categories of words: (i) question words, (ii) words that present a temporal notion of the query, (ii) words that present the subject of tracking, e.g. sleep, walk, and (iv) aggregation words. Our algorithm also uses verb tense for query construction too, which will be explained later.

Figures 1 (a, b, c) plots the frequency of words in categories i, ii and iii, which have been used 10 or more times. To preserve space, we do not plot aggregation words, including: *average*, *miles*, *amount*, *next*, *last*, *more*, *often*, *daily*, etc. Furthermore, users do not always provide aggregation words. Some participants have used a command term to start their query, such as "find", "tell", "give" and "show", which has been shown as "Commands" in Figure 1(a).

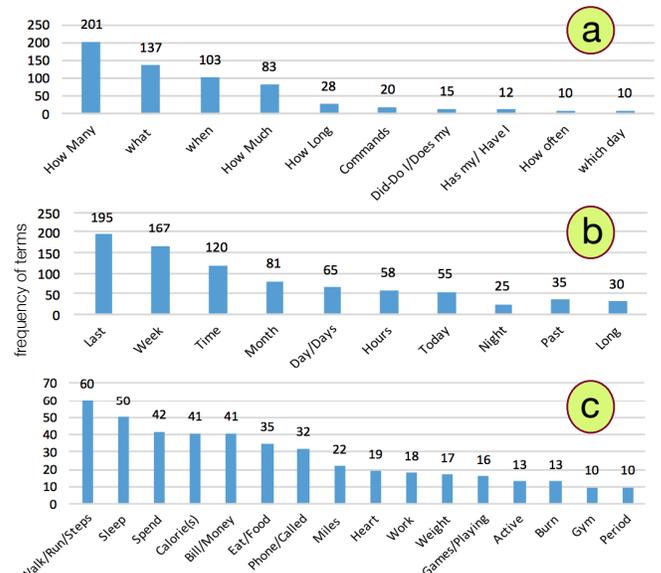

**Figure 1:** Frequency of: (a) questions words, (b) terms that include a notion of time, and (c) terms including the tracking subject, extracted from users' queries.

There are 68 (from 716) queries that do not include the listed question words and have used these command terms. Since these are less than 10% we have neglected to consider them in the implementation. Figure 1(b) includes words that have been used for category (ii). With the exception of 38 queries, all of the other queries included a notion of time, either implicitly or explicitly. In addition to the identified terms, our query parsing algorithm also handles months of the year and days of the week. From our sample, 22 (from 38) queries, which do not include a notion of time, do however include a notion of location (implicit or explicit), such as *"find me a job that matches my qualifications."*, or *"what are the available food deals?"* For these questions, we consider "now" as the notion of time. Figure 1(c) includes the most frequently used words that have been used to describe the subject for tracking.

*3) Tokenization & Lexicon Extraction Algorithm*

Many of existing NLP systems [8,9,10] create a "parse tree" from the given query, which is in text format. Parse tree creation is a computationally complex process but can be used to handle a large variety of texts. However, our approach does not need to cover a wide range of questions. Instead, we focus on QS queries. Moreover, smartwatches can not handle existing parse tree creation algorithm, which usually are computationally complex process. In the evaluation section we describe this in more detail. Here question words (tokenization elements) are limited and known (at least to an extent).

QS queries are usually simple sentences and it is rare that queries include more than one sentences. Therefore, based on the identified word categories, we can parse input queries as a "bag of words". However, the bag of words parsing style has a problem of not considering the "order" of the words. Nevertheless, our term categorization model resolves the need to support order of words. For instance, the following queries have the same semantics; however, they have been written differently. Our parser recognizes them as being similar: *"On average, how often do I eat daily?"*, *"How often, do I eat, on average?"*.[5]

In our model, a query is constituted of five-tuples <$q_w$, $v$, $t$, $s$, $a$>. '$q_w$' for the question word, '$v$' for the verb tense, '$t$' for notion of time, '$s$' for subject(s) of tracking and '$a$' for the aggregation term. There might be queries that do not provide '$a$' and '$t$' explicitly, and thus they should be extracted by the query parser. '$t$' will be either substituted by "now" or based on the tense identified by the last occurrences (maximum of '$t$') of the tracking subject. For instance, *"When did I talk to Sally?"*. By checking the verb tense 'did' the query parser can retrieve the last time that the user has talked to Sally. The evaluation section reports about the impact of '$a$' and '$t$' substitution in more detail. We have found that less than 10% (69 out of 716) of the queries were not retrospective, and were prospective recall e.g. searching for an appointment or doing a prediction e.g. *"When will my next medical check-up be?"* However, there are some queries that perform comparisons. In those queries, there are more than one '$t$'. When the algorithm identifies more than one '$t$', it assumes that a comparison is required and tries to find the nearest aggregation word to the '$t$' (if they are the same). For instance, *"Am I more active this month or last month?"*, or *"Did I sleep more hours on average in March or June?"*.

This version of the algorithm implements the comparison only, it does not support conjunctions terms such as OR, AND, and NOT, because they have not been used in our sample queries.

*C. User Interface*

The tokenization and lexicon extraction model that we have proposed is resource efficient and later we show that the response time of question parsing is very insignificant, that means there is no need to analyze battery use.

However, due its robust input format requirement, the user entry could be error prone. Therefore, similar to the aforementioned NLP works, we rely on user interaction to refine the input. The user interaction should benefit from auxiliary components [8,9] that recommend users what to provide as input into the system.

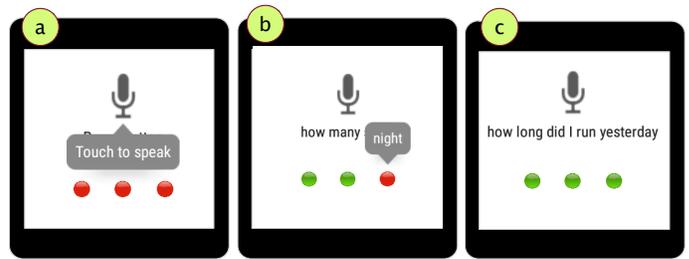

Figure 2 :(a) the user opens the app, (b) a term is missing it has been suggested via tool tip and its button is turned into red, (c) all query terms have been recognized properly.

Nevertheless, considering the miniature-sized screens of smartwatches, our user interface, as it can be seen in Figure 2, recommends only words, and not a complete sentence.

Likewise, it uses three small buttons to notify the user about the missing words in the query. These buttons were assumed as auxiliary controller to assist correcting users' input.

In particular, these buttons are used for '$q_w$', '$t$', '$s$' elements, only words in category i, ii and iii and not '$v$' and '$a$'. If a word, in any of these three categories, has been missed, the related button color will be turned to 'red'. Otherwise the color is green. To our knowledge there are few works that provide UI designs for smartwatches. Despite its simplicity, we believe this interface is among the first works that integrates NLP UI for smartwatches.

III. EVALUATION

We propose four evaluations: query response time, users' accuracy while using the system, and two usability analysis for the smartwatch user interface.

Running the query interface on the device should have a higher response time. To demonstrate this, we have compared our approach with two state-of the-art methods Apache's OpenNLP [16] and Google's TensorFlow, SyntaxNet package [17]. The implementation of OpenNLP and SytaxNet are not light, and they should run outside the Smartwatch. Therefore, we have transferred the data through Bluetooth to the smartphone (OpenNLP), or using the WiFi on the phone to transfer it into the external web server (SyntaxNet). OpenNLP is light enough to run on the smartphone, but not smartwatch. SyntaxNet can not be executed on the smartphone and requires a web server (we used Tomcat Apache Server).

Table 2. shows a comparison of the average response times between our approach and SyntaxNet, OpenNLP for 20 sample queries. As it can be seen, our approach clearly outperforms both methods. For this experiment, we have used a Moto 360 smartwatch (version 2015). The desktop that hosts the NLP server includes a 2.5 GHz Intel Core i5 CPU and 8GB of memory. The

---

[5] We use the underline for "verb words", the red color is for "aggregated words", blue is for "question words", green is for the "notion of time", and magenta is for the "subject(s) of tracking". (Please read this section in color)

smartphone used to transfer the data from the watch to the desktop is Sony Xperia Z5 with 2.0 GHz Quad-core CPU and 3 GB memory. To transfer the data, we have used the embedded Low Power Bluetooth module (BLE) of the smartwatch.

**Table II: Response time comparison between our method (On-Device Parser) and state-of-the-art NLP methods.**

| Method | Host | Response Time (in milliseconds) |
|---|---|---|
| On-Device Parser | Smartwatch | **594** |
| SyntaxNet | Local Server | 3639 |
| OpenNLP | Smartphone | 2271 |

Our experiment uses the WiFi of a local machine. It is notable that in the real-world applications, if we use SyntaxNet the query will be transferred to a cloud and not a local host. Therefore, the parser response time could be more than 3639 milliseconds. Latency is a major challenge for user interaction [18] and results in Table 2 demonstrate that our on-device analysis can resolve the inherit latency of interacting with the smartwatch.

The second evaluation evaluates if the query parser can correctly identify all query elements or not. To analyze each feature, we have selected 10 new test users, 4 males and 6 females, age range from 26 to 35 (mean=29, SD=2.1).

All participants are familiar with both PA and QS systems, we have asked them to issue 6-8 queries using three different settings.

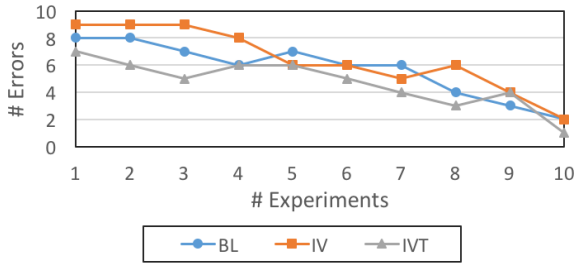

**Figure 3: Errors per query in different text processing modes. The decreasing slope shows that users are learning to use the system and therefore over the time they are getting fewer errors.**

In summary, each user has issued 20-24 queries. A researcher manually checked whether the system could identify all query elements correctly or not. We have used a manual approach, rather than a standard metric, such as "Word Error Rate", because all query elements should be identified correctly and partial correctness of some elements in a sentence is counted as error. The first setting, baseline (BL), is only a bag of words with no additional checking and finds three elements of each query: '$q_w$', '$t$', '$s$'. The second settings checks "the verb tense" (IV) in addition to BL checks. The third settings has checked both "verb tense" and also "time" comparison (IVT).

For the implementation, we have used a third party voice recognition (Google Speech API), and we have neglected voice recognition errors. Voice recognition errors are not in the scope of this paper. Figure 3 shows the number of errors based on the number of experiments. In the beginning, the users made many mistakes. However, as they progressed, they learned the system and performed better. Moreover, IVT performs better than BL and IV. IV also performs slightly better than BL, but not significantly. After the 10[th] query, there are very few errors in creating queries that seems acceptable. We can conclude that by this point, users understand how to construct the query questions correctly and thus the errors have been reduced.

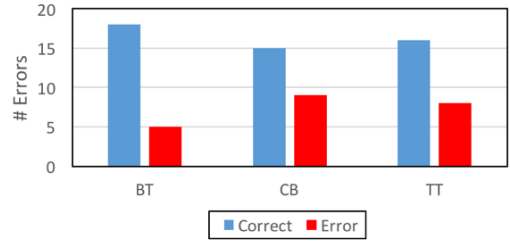

**Figure 4: Error rates based on different UI settings**

The third evaluation focuses on our UI usability and its two main features, i.e. colored buttons and tool tips. In this experiment, the optimal case is enabling both features (BT). We have then enabled the tooltip and disabled the colored buttons (TT). Afterward, we have disabled the tooltip and enabled the colored buttons (CB). As it has been shown in Figure 4, providing the tooltip performs better than the colored buttons, in helping users correctly constructing their queries.

However, as the BT results illustrate, if both tooltip and colored buttons have been used together, we get better results. It is notable that to prevent users learning the system this evaluation has been done in parallel to the previous evaluation. Therefore, the error/correct ratio is rather high in Figure 3. In other words, participants were in learning phase of using the system.

The fourth evaluation is also focuses on the UI usability. For this evaluation we have used Nielesen Heuristics [19]. Nielsen Heuristics evaluates followings: visibility of system status, match between system and the real-world, user control, consistency and standards, error prevention, recognition rather than recall, flexibility and efficiency of use, aesthetic and minimalistic design, help users recognize, diagnose and recover from errors, and finally help and documentations.

We have designed a survey included Nielsen's principles for user interface design. We have adapted those principles in form of questions (in Likert scale) and asked participants to rank the application accordingly. The result of the evaluation shows *Aesthetic and Minimalist Design* received the lowest score (2.9 from 5), but other Nilsen's heuristic factors received satisfactory scores (3.6 – 4.8 from 5). The highest one was

*Recognition rather than recall* that receives the average of 4.8 from 5. Based on participants' feedback the high rank of Recognition rather than recall is because of the use of tooltip in the user interface.

Please consider that the implementation of the information retrieval component is not the scope of this work. Here our focus is on understanding queries and constructing a query interface to identify the required elements of a query from a given sentence.

## IV. CONCLUSION & FUTURE WORK

In this work, we have conducted a user study, through MTurk, to collect a dataset of 716 quantified-self queries that users are willing to issue through their personal assistant systems. We have then presented and evaluated a query interface that includes a user interface and an algorithm for parsing textual quantified-self queries. Our users' accuracy and usability evaluations show that our approach is capable to parse most of the identified quantified-self queries from the survey.

In our future work, we plan to optimize the query interface with subtler term recommendations that are based on the k-nearest neighbor words retrieved from the history of queries. Moreover, we will develop a retrieval component that can search and show behavioral patterns [20] of the smartwatch to the user. This work focused only on query and not information retrieval.